# THE HEAT-BALANCE INTEGRAL METHOD BY A PARABOLIC PROFILE WITH UNSPECIFIED EXPONENT: ANALYSIS AND BENCHMARK EXERCISES

by

## *Jordan HRISTOV*

*The heat-balance integral method of Goodman has been thoroughly analyzed in the case of a parabolic profile with unspecified exponent depending on the boundary condition imposed. That the classical Goodman's boundary conditions defining the time-dependent coefficients of the prescribed temperature profile do not work efficiently at the front of the thermal layers if the specific parabolic profile at issue is employed. Additional constraints based on physical assumption enhance the heat-balance integral method and form a robust algorithm defining the parabola exponent. The method has been compared by results provided by the Veinik's method that is by far different from the Goodman's idea but also assume formation of thermal layer penetrating the heat body. The method has been demonstrated through detailed solutions of 4 1-D heat-conduction problems in Cartesian co-ordinates including a spherical problem (through change of variables) and over-specified boundary condition at the face of the thermal layer.*

## Introduction

In 1958 Goodman [1] has introduced (*in fact the method has bee reported in a conference held in 1957*) the basic heat-balance integral method (HBIM) for solving transient transport problems. The attractive advantage of its approach is the conversion of the governing partial differential equation to an ordinary differential equation by introduction of a prescribed approximation profile satisfying the boundary conditions and integration of the transport equation over the space variable in order to create a heat balance integral. The original quadratic temperature profile is extensively used in literature with both pure fundamental [2, 3] and practically oriented modelling studies [4-12]. The simplicity of the method is still attractive for scientist [13-15] and various modifications [16-18] have been developed.

The core of HBIM is the choice of a profile [19] which should satisfy the conditions at both ends of the thermal penetration depth. The latter, *i. e.* the thermal penetration depth is the physical crux of HBIM (the prescribed profile addresses the mathematical technique of the problem) allowing to present the heating of the medium in an adequate manner thus avoiding the deficiency of the Fourier equation. The common approach is to use a polynomial profile quadratic or cubic to prescribe the temperature profile [2, 20] as they originate from Goodman's classical work [1, 21]. The quadratic profile comes from analytical solutions of problems with simple boundary conditions existing in every textbook on heat transfer. The great advantage of HBIM is the possibility to reduce the partial differential transport equation to an ordinary differential equation with respect to the thermal penetration depth even though strong non-linearity might occur either at the boundaries or as source terms.

In this work we address HBIM with a prescribed temperature profile (see its versions below) expressed generally as:

$$T(x, t) = a + b(1 + cx)^n \qquad (1a)$$

considering in 1-D heating of a slab, with various boundary conditions, namely:

$$\frac{\partial T(x,t)}{\partial t} = \alpha \frac{\partial^2 T(x,t)}{\partial x^2}, \quad 0 \le x \le \delta(t), \quad \delta(0) = 0 \qquad (1b)$$

where $\delta(t)$ is the depth of the thermal penetration layer; the crux of the Goodman's method [1, 21].

In some studies derivatives of eqs. (1), with reduced number of unknown coefficients, have been used. We will briefly comment these works in the light of the present analysis

### Previous work

We refer to some studies employing the prescribed temperature profile of interest rather then to create a survey of its use. The focus is at the methods to define the unknown coefficients with appropriate boundary conditions and specially the exponent since the profile has some specific features at the two ends of the thermal layer.

Mokrushin [22, 23] has proposed the profile:

$$\Theta = \frac{T(x,t) - T_\infty}{T_s - T_\infty} = \left(1 - \frac{x}{\delta}\right)^n \qquad (2)$$

to a method merging the properties of the Goodman's HBIM and the collocation method with a convection boundary condition at $x = 0$.

Similarly, Braga [17, 24] has studied ablation problems with *prescribed heat flux at the boundary* by a profile:

$$\Theta = A\left(\frac{\delta_p - \tilde{x}}{\delta_p}\right)^n \qquad (3a)$$

where $\delta_p = \delta/h_0$ and $\tilde{x} = x/h_0$ are the dimensionless heat penetration depth and distance, respectively, scaled to the slab thickness $h_0$, that gives:

$$\Theta = A\left(1 - \frac{\tilde{x}}{\delta_p}\right)^n = \Theta = A\left(1 - h_0\frac{x}{\delta}\right)^n \qquad (3b)$$

Further, the power-law profile:

$$T = T_s R(\xi), \quad \xi = \frac{x}{\delta} \qquad (4a)$$

with $\delta(t)$ considered as a form parameter has been employed by Pasichnyi *et al.* [25, 26] for slab heating with *prescribed heat flux at the boundary*. The function $R(\xi)$ satisfies the conditions $R(0) = 1$, $R(1) = 0$, and $\partial R/\partial x|_{(x=1)}$ which provide:

$$R(x) = (1 - x)^n \qquad (4b)$$

To this end, Dombrovsky *et al.* [27, 28] have proposed a simplified parabolic profile:

$$\frac{T(r, t) - T_c(t)}{T_S(t) - T_c(t)} = \left(\frac{r}{r_d}\right)^2 \qquad (5a)$$

for heating of evaporating droplet with *Stefan boundary condition at the surface,* namely:

Convection heating $\qquad \lambda \frac{\partial T}{\partial r}\bigg|_{r=r_d} = \alpha_h (T_{gas} - T_s) - \rho_L L \frac{dr_d}{dt} \qquad (5b)$

Since the exponent in expression (5a) is prescribed, *i. e.*, the application of heat-balance integral (HBI) provides an equation about the droplet radius decreasing with time that, in fact, is a modified form of the heat penetration depth represent by the thickness of the evaporated fuel layer $\Delta r = r_d(t = 0) - r_d(t)$.

Further, the profile (2) with $n = 3$ has been used by Antic *et al.* to model the moisture penetration on granular beds [29]. Tarzia [30] has investigated application of the profile (2) in its general form to heat conduction problems with absorption. Besides, the cubic profile with additional smoothing condition (see further in this article) collapses to the profile (2) with $n = 3$ as it has been well exemplified by Ozisik [31, 32]. We will comment these studies in the course of the development of the present study and this initial review gives only information about modifications of the basic profile (1) used so far in the literature. In this context, Spearpoint *et al.* [33, 34] have employed the profile (2) with $n = 1$ to model the pre-burning heating of wood and in the char layer above the virgin layer beyond the ignition point [33, 34].

Similarly the profile $\qquad \Theta = \left(\frac{x}{h_0} - \frac{\delta}{\delta}\right)^2 \qquad (6a)$

with $n = 2$ has been used for ice melting modelling in the post melting period [35].

Generally speaking, the profile (1) with $n = 2$ is quite popular either in its pure form [36, 37] or in a hybrid expressions [39], namely:

$$T(x, t) = T(0, t) \left[ 1 - \frac{x}{\delta(t)} \right]^2 \exp\left( -\frac{x}{\delta(t)} \right) \qquad (6b)$$

within the heat-up period of a laser heated slab.

Additionally to these problems in Cartesian co-ordinates, Lunardini [38] solved a cylindrical analogue of the Neumann problem with the profile with $n = 20$:

$$\frac{T - T_s}{T_s - T_\infty} = \left( \frac{\delta - r}{\delta - r_{00}} \right)^n, \quad r_{00} \le r \le \delta \qquad (7a)$$

The same profile with a hybrid modification has been employed by Harrach [39] for a Stefan problem of a laser vaporizing solid, *i. e.*:

$$\frac{T(x, t) - T(\delta, t)}{T_V - T(\delta, t)} = \left( \frac{\delta - x}{1 - x_s(t)} \right)^2 \exp\left( \frac{x - x_s(t)}{\delta - x_s(t)} \right) \qquad (7b)$$

for $x_s(t) \le x \le \delta(t)$, where $x_s(s)$ is the phase transition line with a temperature $T_V$.

### Motivation for this work

Some of the works briefly reviewed are devoted to specific problems and in most of them the exponents employed are close to $n = 3$ thus in fact repeating the earlier suggestion of Goodman [21] while in the remaining studies $n = 2$ is taken for granted. However, this simplicity does not assure accuracy of the approximation since additional constraint are not commonly used. In the context of these comments, for instance, the profile:

$$T(x) \quad T_i \quad (T_p \quad T_i) \ 1 \quad \frac{x_{char}}{a}^{2} \tag{7c}$$

is accepted as a background for the development of Eurocode 5 (European Standard dealing with timber constructions subjected to fire) [40, 41] as a heat transfer model across wood char layers. Further, the char front temperature $T_p$ is generally assumed to be 300 °C, while $a = 40$ mm is taken as the thermal penetration depth; $x_{char}$ is the depth measured from the char front (in [mm]). This extremely simple approximate profile enables to be compared easily with a simplified analytical solution [42, 43] (and temperature measurements in-depth of wood samples) based on the Carslaw *et al.* error-function solution [44].

These comments give rise to the question concerning a preliminary analysis of the true form of the profile (1) if the complete procedure of the HBIM is applied to it. Preliminary tests indicate that the classical boundary conditions, for instance, commonly employed in HBIM with polynomial profiles (up to that of 4th degree) do not work well in the present case since two of them addressing the temperature derivatives at the front of the thermal layer generate almost identical equations valid for every value of $n$. Hence, a specific problem with the profile (1a) envisaging a definition of additional conditions enabling to define the exponent $n$ exists. This problem is discussed here with two classical examples: prescribed temperature (PT) and prescribed flux (PF) conditions at $x = 0$. These examples enable to evaluate the necessary additional conditions that the profile has to satisfy, mainly based on physical assumptions, rather then from mathematical standpoints, but efficiently defining the value of the exponent $n$. Additionally the method is applied to heat conduction from a sphere to infinite medium in a case allowing simple transformation of the equation by suitable change of variables. As a special problem, the HBIM with the profile at issue is tested with a heat-conduction problem with over-specified boundary conditions at $x = 0$, *i. e.* merging both the PT and PF problems. To our knowledge, there are no detailed studies on the profile (1) oriented to its features and applicability and the discussion developed further in this work highlights specific troubles in its use and how they could be circumvent.

The parabolic profile in the form (2) has been extensively used by Veinik [45] to solve transient heat conduction problems with simple geometries and boundary conditions. This book has appeared in 1959, *i. e.* at almost at the same time as the Goodman article [1] but never has been translated in English. Moreover, the Veinik's approach never has been related to the Goodman's method. To this end some attempts are done in the present article.

At least but not last, bearing in mind the contribution to this special issue of the journal *Thermal Science* dedicated to the 50th anniversary of the Goodman' method [1] the present article is organized as a combination of a lecture and a research report. For the reason of that we address classical problems of transient heating through HBIM solution with the profile (1a) rather than some specific problems with moving boundaries an phase-transition to which the method is commonly applied. The latter group of problems also forms a research trend envisaging an analysis of adequate applications of the profile (1a) but this is beyond the scope of the present work.

The author's experience with teaching since 1981 motivated this approach since in many cases discussions on specific difficulties emerging in the solutions were missing and some findings appeared as *ad hoc* innovations avoiding comments how the problems were fixed. From this standpoint, we may expect that discussions, sometimes missing in research articles with enormous number of equations, will enrich the results developed here.

### Paper scope and outline

The paper, as mentioned at the beginning, addresses application of a parabolic profile with unspecified exponent to the HBI through analytical analysis. The basic steps of the development of this task are outlined as follows:

**Analysis** of the boundary conditions effects on determination of the time-dependent profile coefficients prior to creation of the HBI. The qualitative analysis detects insufficient boundary conditions and impossibility to determine the exponent in the prescribed temperature profile through the rule of the classical HBIM. Further, attempts to define scaling relationships between the exponent $n$ and the thermal layer depth $\delta$ through macroscopic energy balance are performed.

**Exercises** with profiles pertinent to the heat-up stage in two simple cases: isothermal conditions (*prescribed temperature*, *PT*) and *prescribed flux condition* (*PF*) at the face surface of a slab during the heat-up stage. These exercises introduce additional conditions (constraints) through matching the approximate (HBIM) and the exact solution (the Carslaw&Jaeger's solution through the Gauss error-function), *i. e.* (1) equal heat energy accumulated in the thermal layer irrespective of the profile used and the boundary conditions, and (2) equal heat fluxes at ($x = 0$) in the case of PT problems and equal surface temperatures ($x = 0$) with the PF problems. The use of two additional constraints allow an accurate estimation of the exponent $n$ as well as the ratio of the thermal layer depth to the heat diffusion length, *i. e.* $\delta/(\alpha t)^{1/2}$ which becomes a constant depending on the boundary conditions imposed.

### Analysis

Initially, we start with two simplified examples regarding a fixed and know exponent $n$ that simplify the solution but enable to envisage the problem with insufficient conditions (constraints).

### *Solutions with fixed (prescribed) exponent*

Two examples concerning PT and PF condition at the face surface ($x = 0$) will be derived next is order to provide a preliminarily basis for further discussions with unspecified exponent.

– Prescribed temperature at $x = 0$

The boundary conditions imposed on the profile (1) are:

$$T(\delta, t) \quad T_\infty, \quad T(0, t) \quad T_s, \quad \lambda \frac{\partial T}{\partial x}\bigg|_x \quad _\delta \quad 0 \qquad (8a,b,c)$$

that provides

$$a_1 \quad T_\infty; \quad b_1 \quad (T_s \quad T_\infty); \quad c_1 \quad \frac{1}{\delta} \qquad (8d)$$

$$T - T_{\infty} = (T_s - T_{\infty})\left[1 - \left(\frac{x}{\delta}\right)^n\right] \tag{8e}$$

Then, the HBI [1, 21] is:

$$\frac{d}{dt}\int_0^{\delta} T dx - T_{\infty}\delta = \frac{d}{dt}\int_0^{\delta}[a_1 - b_1(1 - c_1 x)^n]dy - T_{\infty}\delta = -\alpha\frac{\partial T}{\partial x}\Big|_{x=0} \tag{9a}$$

with $-\alpha(\partial T/\partial x)\big|_{x=0} = \alpha(n/\delta)(T_s - T_{\infty})$ that follows from the prescribed temperature profile function the depth of the thermal layer is a solution to:

$$\frac{\partial\delta^2}{\partial t} = 2\alpha(n+1)n \Rightarrow \delta_1 = \sqrt{2\alpha(n+1)n}\sqrt{t}, \text{ with } \delta(t=0) = 0 \tag{9b}$$

The temperature profile (7b) then reads:

$$T - T_{\infty} = (T_s - T_{\infty})\left[1 - \frac{x}{\sqrt{\alpha t}\sqrt{2n(n+1)}}\right]^n \tag{10}$$

– Prescribed flux condition $x = 0$

The boundary conditions imposed on the profile:

$$T(\delta, t) = T_{\infty}, \quad -\lambda\frac{\partial T}{\partial x}\Big|_{x=0} = F(t), \quad -\lambda\frac{\partial T}{\partial x}\Big|_{x=\delta} = 0 \tag{11a,b,c}$$

provide

$$a_2 = T_{\infty}; \quad b_2 = \frac{F}{\lambda}\frac{\delta}{n}; \quad c_2 = \frac{1}{\delta} \tag{12a}$$

The general analysis should take into account the time dependency of the flux $F(t)$ but for simplicity of explanation and concentration the efforts on the problem with the exponent $n$, we assume henceforth $F = $ const.

Then, with HBI (eq. 8) and $-\alpha(\partial T/\partial x)\big|_{x=0} = a_2 b_2 (1/\delta)$ we have:

$$\frac{\partial\delta^2}{\partial t} = \alpha(n+1)n \text{ with } \delta(t=0) = 0 \tag{13a}$$

$$\delta_2 = \sqrt{\alpha t}\sqrt{n(n+1)} \tag{13b}$$

Hence the temperature profile is:

$$T - T_{\infty} = \frac{F\delta}{\lambda n}\left[1 - \frac{x}{\sqrt{\alpha t}\sqrt{n(n+1)}}\right]^n \tag{14}$$

In both cases (PT and PF) just exemplified the absence of additional conditions avoids the definition of the exponent $n$. However, with available sets of experiential data the only way to solve practical problems is to apply regression analysis and establish equations similar to the practical ones expressed by eqs. (6) and (7).

### Difficulties emerging at the front of the thermal layer

Generally, with fixed $n$, the number of the required boundary conditions drops to only 3 that clearly distinguish the PT and PF problems. The idea to find boundary conditions defining automatically the exponent $n$ through direct application to the profile (1) generates a problem that is complex and need special attention. In fact we need one condition more since the number of unknowns is 4. Commonly, the fourth condition used with a cubic polynomial profile [1, 20, 21] is $(\partial^2 T/\partial x^2)|_{x\,\delta} = 0$ known as "smoothing condition". In this context, the condition $(\partial T/\partial t)|_{x\,\delta} = \alpha(\partial^2 t/\partial x^2)|_{x\,\delta}$ [5] is equivalent to the "smoothing condition" [2]. Applying the boundary conditions at the thermal layer front $\lambda(\partial T/\partial x)|_{x\,\delta}$ 0 and $(\partial^2 T/\partial x^2)|_{x\,\delta} = 0$ to the power-law profile we have:

$$\lambda\frac{\partial T}{\partial x}\bigg|_{x\,\delta} \quad 0 \qquad 0 \quad \lambda bcn(1 \quad c\delta)^{n\ 1} \tag{15a}$$

$$\frac{\partial^2 T}{\partial x^2}\bigg|_{x\,\delta} \quad 0 \qquad bc^2n(n\ 1)(1 \quad c\delta)^{n\ 2} \quad 0 \tag{15b}$$

These eqs. are almost identical that cast doubts in determination of $b$ and $n$. In contrast to this example, the classical cubic polynomial profile $T(x,t) = a_3 + b_3x + c_3x^2 + d_3x^3$ at $x = \delta$ with (15a, b) generates the equations:

$$\lambda\frac{\partial T}{\partial x}\bigg|_{x\,\delta} \quad 0 \quad b_3 \quad 2c_3\delta \quad 3d_3\delta^2 \tag{16a}$$

$$\frac{\partial^2 T}{\partial x^2}\bigg|_{x\,\delta} \quad 0 \quad 2c_3 \quad 6d_3\delta \tag{16b}$$

but the uncertainty in solution to (15a, b) does not exist.

Bearing in mind that coefficients of profile (1a) are not equal to zero and that the condition $bc^2n(n-1)(1+c\delta)^{n-2} = 0$ is valid in case of $bc^2n(n-1) = 0$ too (that is equivalent to the boundary conditions 15a), the only adequate result provided by conditions (15b) is $n = 1$. In fact, this is the case used by Spearpoint *et al.* [33, 34] to model heat transfer in a char layer of burning wood sample with a flux radiation condition at $x = 0$.

However, as it was demonstrated earlier, the exponent could be any arbitrary value since it is not related to other coefficient of the profile, namely:

$$a_2 \quad T_\infty, \quad b_2 \quad \frac{F(t)}{\lambda}\delta, \quad c_2 \quad \frac{1}{\delta} \tag{17a}$$

The use of $n = 1$ reduces the profile to:

$$T \quad T_\infty \quad \frac{F(t)}{\lambda}\delta\ 1 \quad \frac{x}{\delta} \qquad \frac{T\quad T_\infty}{\frac{F(t)}{\lambda}\delta} \quad 1 \quad \frac{x}{\delta} \qquad \frac{T\quad T_\infty}{\frac{F(t)h_0}{\lambda}} \quad \frac{\delta}{h_0}\ 1 \quad \frac{x}{\delta} \tag{17b}$$

The introduction of the slab thickness $h_0$ makes the profile dimensionless and defines the term $F(t)h_0/\lambda$ as a reference temperature completely determined by PF condition. Similarly, when the thermal layer reaches the end of the slab, *i. e.* $\delta = h_0$ the profile becomes $T(x,t) = a + b(1 - -x/h_0)^n$. The number of the unknowns drops to 3, *i. e. a*, *b*, and *n*, that needs three-boundary conditions only. However, at $x = h_0$ we have $-\lambda(\ T\ \partial x)|_{x\ h_0} = 0$ $0 = \lambda bcn(1-x/h_0)^{n-1}$ that is valid at any $n$. Hence, the problem with insufficient conditions defining the proper exponent remains.

In view of these comments, the common approach is to define preliminarily $a(t)$ and $b(t)$ as $T_\infty$ and $(T_s - T_\infty)$, respectively, [17, 22, 23, 26, 27]. This simplifies the problem and permits the unknown coefficient $c(t)$ and the exponent $n(t)$ to be defined but the problems with the proper determination of $n$ remains since the conditions (15a, b) are satisfied by $c = -1/\delta$ and any $n$ and an additional condition has to be defined.

### Attempts to find conditions defining the exponent $n$

#### – Integral constraints

Generally, we need a condition replacing the smoothing condition and allowing to define the exponent $n$ and we refer to *non-HBIM integral approximations* balancing the energy over the thermal layer.

Let us apply two approximations that will help us to establish the approximate relationship $\delta = f(n)$.

Initially, with the approach of Veinik [45] the heat rate flowing into the thermal layer through the face surface $S$ of the slab at $x = 0$ is:

$$\mathrm{d}Q = \left. -\lambda \frac{\partial T}{\partial x}\right|_{x=0} S \mathrm{d}t \qquad (18a)$$

while the heat accumulated in the thermal layer is:

$$Q = C_\mathrm{p}\rho(T_\mathrm{av} - T_\infty)S\delta = C_\mathrm{p}\rho\left(\frac{1}{\delta}\int_0^\delta T(x,t)\mathrm{d}x - T_\infty\right)S\delta \qquad (18b)$$

That needs the condition:

$$\mathrm{d}Q = \left. -\lambda \frac{\partial T}{\partial x}\right|_{x=0} S \mathrm{d}t = \mathrm{d}[C_\mathrm{p}\rho(T_\mathrm{av} - T_\infty)S\delta] \qquad (19)$$

to be satisfied at $x = 0$. This relationship applied to the two problems considered here yields

PT problem $\qquad \delta = \sqrt{\alpha t}\sqrt{2n(n-1)} \quad \Rightarrow \quad n^2 - n - \frac{\delta^2}{\alpha t} = 0 \qquad (20)$

PF problem $\qquad \delta = \sqrt{\alpha t}\sqrt{n(n-1)} \quad \Rightarrow \quad n^2 - n - \frac{\delta^2}{\alpha t} = 0 \qquad (21)$

that confirms the classical result $\delta \sim (\alpha t)^{1/2}$ [41].

Similarly, after Mokrushin [22, 23] at any $0 \leq x \leq \delta$ the condition:

$$\int_x^\delta \left(\lambda \frac{\partial \Theta}{\partial x} - \frac{\partial Q}{\partial t}\right) \mathrm{d}x = 0 \qquad (22a)$$

should be satisfied.

Here

$$Q = \int_x^\delta \rho C_\mathrm{p}[T(x,t) - T_\infty]\mathrm{d}x \qquad (22b)$$

is the heat accumulated in the slab between cross-sections at $x$ and $x = \delta$.

The assumption $n$ to be a constant over the entire thermal layer and the requirement condition (22a) to be satisfied at $x = \delta/2$ provide (for both PT and PH problems):

$$\frac{\mathrm{d}\delta^2}{\mathrm{d}t} = 8\alpha n(n-1) \quad \Rightarrow \quad \delta = \sqrt{\alpha t}\sqrt{8n(n-1)} \quad \Rightarrow \quad \delta \sim \sqrt{\alpha t} \qquad (23)$$

which in general confirms the estimation (21).

Bearing in mind that the estimations (22a, b) are valid over the entire penetration layer, *i. e.* from $x = 0$ and $x = \delta$, we have:

$$\frac{\partial Q}{\partial t} \approx \rho C_p \frac{(T_s - T_\infty)}{n+1} \delta \approx \lambda \frac{(T_s - T_\infty)}{\delta} \frac{\partial \Theta}{\partial \tilde{x}} \qquad (24)$$

with $\Theta = (T - T_\infty)/(T_s - T_\infty)$ and $\tilde{x} = x/\delta$

Using these estimates, the relationship (24) yields for the PF problem, for example [22, 23], with a convection boundary condition):

$$\frac{d\delta^2}{dt} \approx 8\alpha(n+1)\Phi_0 \qquad \delta \approx \sqrt{\alpha t}\sqrt{8n(n+1)Bi} \qquad (25a)$$

By analogy, if we assume a radiation flux $F$ at $x = 0$ the condition (24) yields:

$$\frac{d\delta^2}{dt} \approx 8\alpha(n+1)\Phi_0 \qquad \delta \approx \sqrt{\alpha t}\sqrt{8n(n+1)\Phi_0} \qquad (25b)$$

Here the dimensionless numbers $\Phi_0 = Fh_0/\lambda(T_s - T_\infty) = \Theta/\tilde{x}$ and $Bi = \alpha_h h_0/\lambda = \Theta/\tilde{x}$ comes from non-dimensionalization of the flux and convection boundary conditions at $x = 0$, respectively.

– **Optimal values of n**

The optimal value of $n$ used in previous works of Mokrushin [22, 23], for instance, was found to be $n = 2$ at $Bi = 0$ and $n = 4$ at $Bi \rightarrow \infty$. That is the average value is about $n_{av} = 3$. Further, Braga [17, 24] got $n = \pi/(4 - \pi) \approx 3.65979$ with a flux boundary condition though a procedure balancing the heat energy accumulated in the body provided by HBIM solution with that calculated through Gauss error-function [44] assumed as exact one. In contrast to these attempts to find a approximate analytical relationship defining the exponent $n$, Pasichnyi *et al.* [25, 26] directly use $n = 2$ with explanation that the solution is more physical in nature than mathematical one. Similarly Ho *et al.* [35] and Zhu *et al.* [36] used $n = 2$ for heat conduction in rectangular and cylindrical geometry, respectively, with flux condition at the face surface.

In fact, these comments confirm that all the results oscillate around $n = 3$ as suggested by an example provided by Goodman [21]. However, the solution of Lunardini [39] in cylindrical coordinates with the profile (7a) teaches that the accuracy of the solution with respect to the exact result of Carslaw *et al.* [42] above $n = 20$ is sufficient, so it could be assumed as an extreme.

In the context of the results developed in the previous section in both final solutions there is a term $n(n + 1)$ that make the problem strongly non-linear. To circumvent the problem we credit the work of Rosenband *et al.* [46] where the simplification $n/(n + 1) \approx 1$ was suggested. That is, a simple test with the data of Mokrushin [22, 23] give $0.66 \approx n/(n+1) \approx 0.8$ while the value of Braga *et al.* [17]. *i. e.*, $n \approx 3.65979$ yields $n/(n + 1) = 0.7853$. Further, with $n = 3$ we have $n/(n + 1) = 0.75$. The assumption $n/(n + 1) \approx 1$ is a very strong simplification, but bearing in mind that:

$$\frac{n}{n+1} \approx 1 \qquad \frac{n^2}{n(n+1)} \approx 1 \qquad n(n+1) \approx n^2 \qquad (26)$$

the two relationships $\delta = f(n)$ can be read (21a, b) as:

$$\delta \approx \sqrt{\alpha t}\sqrt{2n(n+1)} \qquad n\sqrt{2}\sqrt{\alpha t} \qquad (27a)$$

$$\delta = \sqrt{\alpha t}\sqrt{8n(n-1)} \qquad \delta = \sqrt{\alpha t}\sqrt{8n^2} \qquad \delta = n2\sqrt{2}\sqrt{\alpha t} \tag{27b}$$

$$\delta = n2\sqrt{2}\sqrt{\alpha t}\sqrt{\Phi_0} \quad \text{and} \quad \delta = n2\sqrt{2}\sqrt{\alpha t}\sqrt{\text{Bi}} \tag{28c,d}$$

Hence, in general from (25a, b) we have:

$$n = \frac{\delta}{\sqrt{\alpha t}}\frac{1}{\sqrt{\Phi_0}} \quad \text{or} \quad n = \frac{\delta}{\sqrt{\alpha t}}\frac{1}{\sqrt{\text{Bi}}} \tag{29a,b}$$

That may be expressed as:

$$n = p_{\text{F}}\frac{\delta}{\sqrt{\alpha t}}\frac{1}{\sqrt{\Phi_0}} \quad \text{or} \quad n = p_{\text{C}}\frac{\delta}{\sqrt{\alpha t}}\frac{1}{\sqrt{\text{Bi}}} \tag{30a,b}$$

with tuning coefficients $p_{\text{C}}$ and $p_{\text{H}}$ that have to be determined through comparison with the exact solution or experimental data (subscripts $C$ and $F$ means *convective* and *flux* condition) bearing in mind that Bi number and $\Phi_0$ are constants.

These attempts clearly indicate that if the exponent is chosen to be a constant then the ratio $\delta/(\alpha t)^{1/2}$ has to be a constant too since it depends on the boundary conditions only. Therefore, the problems in proper determination of the exponent $n$ still exist since they are related to the diffusion (thermal) length scale $(\alpha t)^{1/2}$. In this context, the estimations (21), (23), and (25a, b) give an idea that the ratio $\delta/(\alpha t)^{1/2}$ might be a constant (still unknown) thus simplifying the determination of $n$. However, with this assumption we define one unknown, $\delta/(\alpha t)^{1/2}$, more that requires a definition of two constraints instead one looked for defining of $n$ as it is assumed in [17, 22-24].

### Additional constraints defining the exponent

The mathematical exercises performed in this section address problems discussed earlier, *i. e.* with PT and PF, but additional constraints employing physical integral assumptions and Gauss error-function based solutions [41] are employed to define both the term $\delta/(\alpha t)^{1/2}$ and $n$. Therefore, all examples further address the heat-up stage when both $\delta$ and $n$ are unknown.

**Exercise 1** with PT problem (isothermal condition at $x = 0$)

With $T = T_{\text{s}}$ at $t = 0$ and $T(x, 0) = T_{\infty}$ at $x = 0$ the classical solution through the Gauss error-function [42] is:

$$\frac{T(x,t) - T_{\text{s}}}{T_{\infty} - T_{\text{s}}} = \text{erf}\left(\frac{x}{2\sqrt{\alpha t}}\right) \quad \text{with} \quad \text{erf}\left(\frac{x}{2\sqrt{\alpha t}}\right) = \frac{2}{\sqrt{\pi}}\int_0^{\frac{x}{2\sqrt{\alpha t}}} e^{-\eta^2}d\eta \tag{31a,b}$$

Two physically based assumptions (constraints) could be formulated, namely:
- the flux at $x = 0$ provided by the approximate ($q_{\text{a}}$) and the exact solution ($q_{\text{e}}$) should be equal, *i. e.* $q_{\text{a}}(x = 0) = q_{\text{e}}(x = 0)$, that is the *1st constraint*, and
- the heat accumulated in the body up to a time $t = t$ corresponding to certain $\delta(t)$ should equal too, that is $Q_{\text{a}}(x)\big|_0^\delta = Q_{\text{e}}(t)\big|_0^t$ irrespective of the type of temperature profile used. This is the *2nd constraint*.

With the first condition we have:

$$q_{\text{e}}(x = 0) = \frac{\lambda(T_{\text{s}} - T_{\infty})}{\sqrt{\pi}\sqrt{\alpha t}} \quad \text{and} \quad q_{\text{a}}(x = 0) = \frac{\lambda}{\delta}(T_s - T_{\infty})n \tag{31a,b}$$

The condition $q_a(x=0) = q_e(x=0)$ yields:

$$\frac{\delta}{\sqrt{\alpha t}} = n\sqrt{\pi} \qquad (31c)$$

The heat accumulated in the body calculated through the approximate and the exact profiles are:

$$Q_a(t)\big|_0^t = \rho C_p (T_s - T_\infty)\frac{\delta}{n+1} \qquad (32a)$$

$$Q_e(t)\big|_0^t = \int_0^t q_e(t)\mathrm{d}t = \frac{\lambda(T_s - T_\infty)}{\sqrt{\pi\alpha}}\int_0^t \frac{1}{\sqrt{t}}\mathrm{d}t = 2\lambda(T_s - T_\infty)\sqrt{\frac{t}{\pi\alpha}} \qquad (32b)$$

The condition $Q_a(t)\big|_0^\delta = Q_e(t)\big|_0^t$ yields:

$$\frac{\delta}{\sqrt{\alpha t}} = 2\frac{2n+1}{\sqrt{\pi}} \qquad (32c)$$

Combining the results of both constraints, *i. e.* from (31c) and (32c) we have:

$$\frac{\delta}{\sqrt{\alpha t}} = n\sqrt{\pi} = 2\frac{2n+1}{\sqrt{\pi}} \qquad (33a)$$

That results in:

$$n = \frac{2}{\pi - 2} = 1.75 \qquad (33b)$$

This result confirms that of Braga *et al.* [17] albeit only the second constraint was used, *i. e.*, $Q_a(t)\big|_0^\delta = Q_e(t)\big|_0^t$ and the integration was performed from 0 to ∞ with respect to the similarity variable $\eta = x/(\alpha t)^{1/2}$. The last results (31c, 33c, and 33b) confirm the suggestion done in advance that $\delta/(\alpha t)^{1/2} = $ const. However, this constant, as well as the value of the exponent $n$ depend on the boundary conditions applied at $x = 0$, and for the PT problem (from 31c) is $\delta/(\alpha t)^{1/2} = 3.10$. This is a reasonable result since the *thermal layer depth* $\delta$ and the *heat penetration depth* $(\alpha t)^{1/2}$ have different physical meanings and always $\delta/(\alpha t)^{1/2} > 1$. In this context, in order to avoid further misinterpretations, we have to mention that HBIM and the exact solution use different length scales for non-dimensionalization of $x$, *i. e.* at $\delta$ and $(\alpha t)^{1/2}$, respectively.

***Exercise 2*** with PF (flux radiation condition at $x = 0$) problem

The exact solution [44] is:

$$T_e(x,t) = \frac{2F}{\lambda}\sqrt{\alpha t}\;i\mathrm{erfc}\left(\frac{x}{2\sqrt{\alpha t}}\right) = \frac{2F}{\lambda}\sqrt{\alpha t}\left[\frac{1}{\sqrt{\pi}}\exp\left(-\frac{x^2}{4\alpha t}\right) - \frac{x}{2\sqrt{\alpha t}}\left(1-\mathrm{erf}\frac{x}{2\sqrt{\alpha t}}\right)\right] \qquad (34)$$

$i\mathrm{erfc}[x/2(\alpha t)^{1/2}] = \int_0^x \mathrm{erfc}(s)\mathrm{d}s$ is the integral of the complimentary error-function $\mathrm{erfc}(h) = 1 - \mathrm{erf}(h)$ with a similarity variable $\eta = x/(\alpha t)^{1/2}$.

The flux condition at $x = 0$ is satisfied by both profiles, that is $q_a(x=0) = F(t)$ and $q_e(x=0) = F(t)$, so additional condition at the face surface is required. In the PF case, we have to ask $T_a(0,t) = T_e(0,t)$.

Hence with:

$$T_e(0,t) - T_\infty = \frac{2F}{\lambda\sqrt{\pi}}\sqrt{\alpha t} \quad \text{and} \quad T_a(0,t) - T_\infty = \frac{F\delta}{\lambda n} \qquad (35a, b)$$

we have

$$\frac{2F}{\lambda\sqrt{\pi}}\sqrt{\alpha t} = \frac{F\delta}{\lambda n} \Rightarrow \frac{\delta}{\sqrt{\alpha t}} = \frac{2n}{\sqrt{\pi}} \qquad (36a, b)$$

Since $\delta = (\alpha t)^{1/2}[n(n+1)]^{1/2}$ (see 13b), we have:

$$\sqrt{n(n+1)} \ \frac{2n}{\sqrt{\pi}} = \frac{n+1}{n} \ \frac{4}{\pi} \quad n = \frac{\pi}{4-\pi} = 3.65 \tag{37}$$

This approximation confirms the result of Braga *et al.* [17] obtained through the condition $Q_a(t) = Q_e(t)$ similar to the operations performed next.

Further, the application of the second constraint, *i. e.* $Q_a(t)\big|_0^\delta = Q_e(t)\big|_0^t$ yields:

$$\frac{F\delta}{\lambda}\sqrt{\alpha t} \ \frac{\sqrt{n(n+1)}}{n+1} = \frac{F\delta}{\lambda} \ \frac{\sqrt{\pi}}{2}\alpha t \tag{38}$$

In fact we have $Q_e(t)\big|_0^t = Q_e(t)\big|_0^\delta$ since beyond $x = 0$ the medium is thermally undisturbed. From (38) we read:

$$\sqrt{\frac{n}{n+1}} = \frac{\sqrt{\pi}}{2} \quad n = \frac{\pi}{4-\pi} = 3.65 \tag{39}$$

This is the same result obtained with the condition $T_a(0,t) = T_e(0,t)$. It confirms the result of Braga *et al.* [17] obtained through $Q_a = Q_e$ but their integration procedure is slightly different from that developed here and repeats the classical one of Carslaw *et al.* [44]. Similarly to the analysis carried out with the PT problem, with (39) we have from (36b) $\delta/(\alpha t)^{1/2} = 2n/\pi^{1/2} = 4.23$.

### Brief comments on the results provided by exercises 1 and 2

The two analytical exercises performed clearly indicate that two additional physically based constraints are needed to determine the exponent *n* in a definitive manner. Additionally to the requirement the heat accumulated in the thermal layer to be the same irrespective of the type of temperature profile used for this, one more condition (constraint) has to be defined for proper determination of the unknown exponent. Generally, the value of *n* depends on the boundary condition applied at $x = 0$. In the case of PT problem we need two conditions, while with the PF problem only the overall heat balances constraint is enough to determine the exponent *n*. On the other hand, additional qualitative result was obtained through these exercises, *i. e.* $\delta/(\alpha t)^{1/2}$ is constant depending on the boundary condition prescribed at $x = 0$. In fact with both exercises we solved sets of algebraic equations with two unknowns, namely $\delta/(\alpha t)^{1/2}$ and *n* physically (intimately) related through the conservation of the energy over the thermal layer and the conditions imposed at $x = 0$ to HBIM and the exact profiles. The fact is not strange and the earlier results of Mioura [47] are close to the values of $\delta/(\alpha t)^{1/2}$ developed here.

***Exercise 3.*** A problem with simultaneously defined PT and PH *conditions at* $x = 0$.

Such a case emerges when the face surface $x = 0$ *is a phase change isotherm* with evaporation (or vaporization) of a liquid burning in pool [12, 48]. In this case the PH condition might be either the convection one (evaporation) or the Stefan boundary condition (SBC). Generally, these situations generates 1-D heat conduction problems with moving boundary problems but as an initial step let us see how the boundary conditions define the coefficients of the prescribed temperature profile. The coefficients *a*, *b*, *c* and the exponent *n* depend on the boundary conditions imposed on the profile, namely:

$$\lambda\frac{\partial T}{\partial x}\bigg|_{x=0} = F(t), \quad T(0,t) = T_s, \quad \lambda\frac{\partial T}{\partial x}\bigg|_{x=\delta} = 0, \quad \text{and} \quad T(\delta, t) = T_\infty \text{ (40a, b, c, d)}$$

The condition (40a) defines $T_s$, a specified reference temperature, commonly the temperature of phase transition. Besides, the $F(t)$ is a general description of the flux at $x = 0$. The application of these *over-specified* conditions to the profile (1) gives:

$$a = T_\infty, \quad b = T_s - T_\infty, \quad c = \frac{1}{\delta}, \quad n = \frac{F(t)}{\lambda(T_s - T_\infty)}\delta \qquad (41a, b, c, d)$$

and the dimensionless profile (1) as:

$$\Theta = \frac{T(x,t) - T_\infty}{T_s - T_\infty} = \left(1 - \frac{x}{\delta}\right)^{\Phi\delta} \qquad (42)$$

In (42) $\Phi = F(t)/\lambda(T_s - T_\infty)$ has a dimension of 1/m, so the exponent $n$ is dimensionless. Then the HBI (eq. 8) provides (with $\delta = 0$ at $t = 0$):

$$\frac{d}{dt}\left[\frac{\delta}{1 + \Phi\delta}\right] = a\frac{F(t)}{\lambda(T_s - T_\infty)} = a\Phi \qquad (43)$$

$$\delta = \frac{1}{\Phi}[\exp(a\Phi^2 t) - 1] \qquad \frac{\delta}{\sqrt{\alpha t}} = \frac{1}{\sqrt{Fo}}\frac{1}{\Phi_0}[\exp(a\Phi^2 t) - 1] \qquad (44a,b)$$

with $Fo = \alpha t/h_0^2$ and $\Phi_0 = F(t)h_0/\lambda(T_s - T_\infty)$ the exponent $n$ is:

$$n = \frac{F(t)}{\lambda(T_s - T_\infty)}\delta = \Phi\delta = [\exp(a\Phi^2 t) - 1] \quad n = \exp(Fo\Phi_0^2) - 1 \qquad (45)$$

Then the temperature profile is:

$$\frac{T(x,t) - T_\infty}{T_s - T_\infty} = \left(1 - \Phi\frac{x}{\exp(a\Phi^2 t) - 1}\right)^{\exp(a\Phi^2 t) - 1} = \left(1 - \frac{x}{h_0}\Phi_0\frac{1}{\exp(Fo\Phi_0^2) - 1}\right)^{\exp(a\Phi_0^2) - 1} \qquad (46a, b)$$

Since $n = F(t)/\lambda(T_s - T_\infty)\delta = \exp(Fo\Phi_0^2) - 1$ is a time-dependent function it is interesting to define its upper limit. Recall that $n = \Phi\delta$, that is $n = \Phi_0(\delta/h_0)$ and at the moment when $\delta = = h_0$ the upper extreme is $n_{max} = \Phi_0$ reached at the end of the heat-up stage, *i. e.* at HBI-$t_h = = (h_0^2/\alpha)(1/\Phi_0^2)]\ln(1/\Phi_0)$. Hence, with the specific problem discussed here $n(\delta) = n(t)$ varies within the range $0 < n(t) < \Phi_0$. Beyond this point, *i. e.* at $t > t_h$ the procedure with HBIM should be repeated but with reduced number of unknown coefficients since $c = -1/h_0$ and $n = $ const. do not vary with time and the problem goes down to that solved in example 2. Further, constraints such those employed in examples 1 and 2 has to be applied to determine the exact value of $n$.

At the end of this example we give short comments on the dimensionless variables emerging through the solution developed. Because the problem address a slab of finite thickness $h_0$ the flux defined length scale $1/\Phi$ was expressed in a dimensionless form as $\Phi h_0 = \Phi_0$. In fact, $\Phi_0 = F(t)h_0/\lambda(T_s - T_\infty) = N_0$ is the so-called "radiation-conduction" number [48] which is analogous the Biot number in case of a convection boundary condition. The dimensionless depth of the thermal layer is expressed by (44b) as $\delta/h_0 = (1/N_0)[\exp(FoN_0^2) - 1]$ with a Fourier number $Fo = Fo_c = at/h_0^2$ defined through the slab thickness as a length scale. The relationship between $Fo_c$ and the flux defined Fourier number $Fo_f = a\Phi^2 t$ with $1/\Phi$ as a length scale (44a, b), (45), and (46a, b), is $Fo_c = Fo_f/N_0^2$. Hence, the short-term ($Fo_c \ll 1$) expression of $\delta/h_0$ (at $FoN_0^2 \ll 1$:

$$\Theta = \left(1 - \frac{x}{h_0}\frac{1}{N_0 Fo_c}\right)^{N_0^2 Fo_c} \qquad (47)$$

***Exercise 4.*** Spherical geometry (hot sphere immersed in an infinite medium)

Let us consider a hot sphere of a radius $r_0$ with isothermal surface, *i. e.* $T = T_s$ at $r = r_0$ immersed at $t = 0$ a in medium with uniform temperature $T_\infty$. For simplicity, we assume that the temperature field is $T(r, t)$ only and the heat transfer equation is (ref. 33, p.158):

$$\frac{1}{\alpha}\frac{\partial T}{\partial t} = \frac{\partial^2 T}{\partial r^2} + \frac{2}{r}\frac{\partial T}{\partial r} \qquad (48a)$$

where $r = 0$ is at the sphere center.

With a new dependent variable $U(r, t) = rT(r, t)$ eq. (48) can be written as:

$$\frac{\partial U}{\partial t} = \alpha\frac{\partial^2 U}{\partial r^2} \qquad (48b)$$

which is now one-dimensional, time-dependent heat conduction in the rectangular co-ordinates (example 1). The following solution use the techniques of example 1 bearing in mind the transformation of variables just done.

From the definition of $U$ and the boundary conditions, one obtain ($\partial U/\partial r)\big|_{r,\delta} = T(d, t) + \delta(\partial T/\partial r)\big|_{r,\delta} = T_\infty$, then equation above is written for $T_\infty = 0$. Initially, the assuming the prescribed profile in the form (1a) we have:

$$U(r,t) = \beta_0 + \beta_1(1 + \beta_2 r)^n \qquad (49)$$

with boundary conditions:

$$U(r_0,t) = U_s, \quad U(\delta,t) = U_\infty, \quad \frac{\partial U}{\partial r}\bigg|_{r=\delta} = 0 \qquad (50)$$

Further, applying the HBI conditions we have:

$$U_s = \beta_0 + \beta_1(1 + \beta_2 r_0)^n \qquad (51a)$$

$$U_\infty = \beta_0 + \beta_1(1 + \beta_2\delta)^n \qquad (51b)$$

$$0 = n\beta_1\beta_2(1 + \beta_2\delta)^{n-1} \qquad (51c)$$

The coefficients are:

$$\beta_0 = U_\infty, \quad \beta_1 = \frac{(U_s - U_\infty)}{\left(1 - \dfrac{r_0}{\delta}\right)^n}, \quad \beta_2 = -\frac{1}{\delta} \text{ and any } n \qquad (52)$$

The coefficient $\beta_2$ matches that of example 1 when $r_0 = 0$, *i. e.* the heat is generated by a point source at $r = r_0$. Equation (51c) is valid for any $n$ and the only useful solution is $\beta_2 = -1/\delta$. This calls for additional constraint defining the exponent $n$. Hence, the profile (49) becomes:

$$U(r,t) = U_\infty + \frac{U_s - U_\infty}{\left(1 - \dfrac{r_0}{\delta}\right)^n}\left(1 - \frac{r}{\delta}\right)^n \quad U(r,t) = U_\infty + (U_s - U_\infty)\left(\frac{\delta - r}{\delta - r_0}\right)^n \qquad (53a,b)$$

The backward transformation $U(r, t) \to T(r, t)$ gives the following form of the profile (1):

$$T(r,t) = T_\infty + (T_s - T_\infty)\left(\frac{\delta - r}{\delta - r_0}\right)^n \qquad (53c)$$

The form (53c) matches the profile of Lunardini [38] used in a cylindrical problem with a $T(r, t)$ field (no $z$ component). However, the further steps of the solution continue with the variable $U(r, t)$ since it satisfies eq. (48b) over which the HBI method is classically applicable.

With the profile (53) the HBI of eq. (48b), namely:

$$\frac{d}{dt}\int_{r_0}^{\delta} U \, dr - U_\infty \delta + \alpha \left.\frac{\partial U}{\partial r}\right|_{r-r_0} - \frac{(U_s - U_\infty)}{n+1}\frac{d\delta}{dt} - \alpha\frac{n}{\delta - r_0}(U_s - U_\infty) \qquad (54a, b)$$

yields

$$\frac{d(\delta - r_0)^2}{dt} - 2\alpha n(n+1) \quad \text{with} \quad \delta - r_0 - 0 \quad \text{at} \quad t - 0 \qquad (55)$$

The thermal layer (shell around the sphere) thickness becomes:

$$(\delta - r_0)^2 - 2\alpha n(n+1)t \qquad (\delta - r_0) - \sqrt{\alpha t}\sqrt{2n(n+1)} \qquad (56a, b)$$

From (56b) we have:

$$\frac{(\delta - r_0)}{\sqrt{\alpha t}} - \sqrt{2n(n+1)} \qquad (57)$$

With the exact solution of (48b) from Carslaw *et al.* [44]

$$U(r, t) - \text{erf}\left(\frac{r}{2\sqrt{\alpha t}}\right) \qquad (58)$$

the constrain $q_a = q_e$ at $r = r_0$ gives:

$$\lambda\frac{n(U_s - U_\infty)}{\delta - r_0} - \lambda\frac{U_s - U_\infty}{\sqrt{\pi}\sqrt{\alpha t}} \qquad (58a)$$

From (58a) we have:

$$\frac{\delta - r_0}{\sqrt{\alpha t}} - n\sqrt{\pi} \qquad (58b)$$

Further, from (57) and (58b) we have:

$$\sqrt{2n(n+1)} - n\sqrt{\pi} \quad n - \frac{2}{\pi - 2} - 1.75 \qquad (59)$$

which matches the result obtained in exercise 1.

Additionally, the constrain $Q_a|_0^\delta - Q_e|_0^t$ results in:

$$\rho C_p(U_s - U_\infty)\frac{\delta - r_0}{n+1} - 2\lambda(U_s - U_\infty)\sqrt{\frac{t}{\pi\alpha}} \qquad (60a)$$

and

$$\frac{\delta - r_0}{\sqrt{\alpha t}} - \frac{2(n+1)}{\sqrt{\pi}} \qquad (60b)$$

Combining (58b) and (60b) we have $n = 2/(\pi-2) - 1.75$ which is the same as obtained by combination of (57) and (58b). In fact, in the case of isothermal boundary condition at the face of the thermal layer the heat-balance integral is equivalent to $Q_a|_0^\delta$ and its time-derivative balances the flux at $x = 0$ which is the crux of the Goodman method [1, 21]. With $n = 1.75$ from (57) we have $(\delta - r_0)/(\alpha t)^{1/2} - 3.10$ like in example 1.

Therefore the temperature profile in the medium surrounding the sphere is:

$$T(r, t) - T_\infty - (T_s - T_\infty)\left(\frac{\Delta_r}{\Delta_0}\right)^n \qquad (61)$$

with $\Delta_r = \delta(t) - r$ and $\Delta_0 = \delta(t) - r_0$.

**Comments on the accuracy of the profile under discussion**

Commonly the accuracy of the approximate solution is tested through direct comparison with results generate by exact solutions of the problem solved if available. Otherwise, the only way is compare the HBIM solution with experiential data. The second approach is commonly employed in the literature [16-18, 20, 22-26]. The approximate profile at issue practically has not been tested for accuracy neither with exact solutions nor experimental data. This is not the principal problem even though with appropriate experimental data this problem could be fixed. The problem is of primary interest since the profile (1a) is equally employed to solve either prescribed temperature or prescribed flux problems with $n = 2$ for example, while the present analysis and the works of Braga *et al.* (2005) and Braga (2006) clearly indicate that the exponents are 1.75 and 3.75 approximately. Here we will follow the approach of Langford [2] to identify both the accuracy of the profile (1a) and additional conditions on the exponent *n*. Thus, if an exact solution exists, as in the test problems developed in the exercises here, the function $E(t)$ [2] could be created:

$$E(t) = \int_0^{\delta(t)} \left[ \frac{\partial T_a(x,t)}{\partial x^2} - \frac{1}{\alpha} \frac{\partial T_a(x,t)}{\partial t} \right]^2 dx \geq 0 \qquad (62)$$

If $T_a(x, t)$ matches the exact solution $T_e(x, t)$, then $E(t)$ would be identically zero by eq. (1b). The effect of the square under the integral is prevent algebraic canceling of errors of opposite sign [2] and to identify the space regions where $T_a(x, t)$ does not strongly satisfy the conduction eq. (1b). According to Langford [2] if $E_n(x, t)$ is a result corresponding to $T_a(x, t)$ and $\delta(t)$, then it can be expressed as:

$$E_n(t) = e_n [\lambda(T_s - T_i)]^2 \frac{1}{\sqrt{\alpha t}}, \quad e_n = \text{const.} \qquad (63)$$

The lower the value of $E_n(x, t)$ or of $e_n$ the more accurate the solution is expected to be. Hence, with the PT problem, for example, we have:

$$E_n = \int_0^{\delta} \left( \frac{T_a - T_e}{T_s - T_\infty} \right)^2 dx = \Delta \int_0^1 \left[ 1 - \left( \frac{X}{\sqrt{2n(n-1)}} \right)^n - \text{erf}(X) \right]^2 dX \qquad (64a)$$

with

$$0 \leq X \leq \frac{x}{\sqrt{\alpha t}} \leq \frac{\delta}{\sqrt{\alpha t}} \quad \text{and} \quad \Delta = \frac{\delta}{\sqrt{\alpha t}} \qquad (64b)$$

The solution performed in exercise 1 reveals that with $n = 1.75$ we have $X(x = \delta) = \delta/(\alpha t)^{1/2}$ (In fact at $x = \delta$ we have $\delta/(\alpha t)^{1/2} = [2n(n+1)]^{1/2}$ – see eq. (10).

Hence, the integral defining $E_n$ with $n = 1.75$ is (done by integration with a Maple 9.5 solver):

$$E_{(n=1.75)} = \int_0^{3.1} [(1 - 0.332X)^{1.75} - \text{erf}(X)]^2 dX = 1.64674 \qquad (65)$$

Similarly, with other values of the exponent met in this text we have:

$$-n = 2 \qquad\qquad E_{(n=2)} = \int_0^{3.46} [(1 - 0.288X)^2 - \text{erf}(X)]^2 dX = 1.91332 \qquad (66a)$$

$$-n = 3 \qquad\qquad E_{(n=3)} = \int_0^{4.89} [(1 - 0.204X)^3 - \text{erf}(X)]^2 dX = 3.207569 \qquad (66b)$$

$$- n = 3.65 \qquad E_{(n=3.75)} \int_0^{5.96} [(1 - 0.167X)^{3.75} - \operatorname{erf}(X)]^2 \mathrm{d}X = 4.20960 \qquad (66c)$$

$$- n = 4 \qquad E_{(n=4)} \int_0^{6.32} [(1 - 0.158X)^4 - \operatorname{erf}(X)]^2 \mathrm{d}X = 4.5567 \qquad (66d)$$

and with an extreme value of the exponent as well:

$$- n = 20 \qquad E_{(n=20)} \int_0^{28.98} [(1 - 0.034X)^{20} - \operatorname{erf}(X)]^2 \mathrm{d}X = 26.9550 \qquad (66e)$$

The results do not look strange since the minimal value of $E_n$ ($n = 1.75$) is obtained with an approximate profile that even though matches the conditions of the HBI is constrained additionally by two requirements $i.$ $e.$ $q_a(x = 0) = q_e$ $(x = 0)$ and $Q_a|_0^\delta = Q_e|_0^t$. The first one needs $T_a / x = T_e / x$ at $x = 0$ whilst the second one (an integration over the thermal layer) is a requirement similar to that imposed by $E_n$, namely:

$$\delta Q = Q_a|_0^\delta - Q_e|_0^t = \int_0^\Delta \rho C_p \left[ 1 - \frac{X}{\sqrt{2n(n-1)}} \right]^n - \operatorname{erf}(X) \, \mathrm{d}X = 0 \qquad (67)$$

Thus, it is expected that the minimum of $\delta Q$ should be satisfied at conditions close to that where we the minimum of $E_n$ exists. To this end, we have to mention that $n = 1.75$ is established in the Veinik's works by a simple comparison the exact solution and data provided by the approximate solution without clear algorithm defining the optimal exponent value and criteria leading to minimum error.

Therefore, the additional constraints used to define the exponent of the prescribed profile automatically provide a procedure similar to that just performed with PF problem (details are not documented here) and confirms the physically adequate exponent $n = 3.65$.

Further, a reasonable question is: is it possible to use integer exponents in calculations as more convenient procedures, $i.$ $e.$, to replace 1.75 by 2 and 3.65 by 4 in the PT and PF problems, respectively? Simple calculations give $[E_n (n = 2) - E_n (1.75)]/E_n (1.75) = 0.1618$ that implies an increase in approximately 17% with respect to the correctly defined equivalent temperature profile. In this context, the principle call coming from this study is to find in many cases emerging in practice and requiring solutions through HBIM adequate additional constraints. The additional constraints imposed at $x = 0$ in the solved examples are easily applicable since most of the exact solutions available in literature [31, 32, 44] are valid at small $x$ and gives straightforward answers at $x = 0$ either for the heat flux or the isothermal boundary condition. The second constraint depends on precision of the integration procedure of the exact solution that sometimes is easily to perform but occasionally not. In this case computer algebras like Maple or Mathematica could facilitate the solution.

## Brief discussion and results outline

The analysis and the supporting exercises went through many steps towards a well defined standpoint concerning specification of the prescribed parabolic profile and especially of its exponent. The analysis and the exercises deal with simple problems and the only goal is to define the problem at issue in an academic manner. This was especially done since in many case referred in this work the exponent of the approximating parabola is prescribed either as 3 or 2. The latter avoids many problems in the determination of the coefficients of the profile as commented at the beginning of this article. However, this simplicity does not assure accuracy of the approxi-mation since additional constraints are not commonly used. The only way is to test the approximated profile with either experimental data or exact solutions if they exists.

The test of accuracy of the correct choice of the approximate profile through the approach of Langford [2] reveal that the Langford's condition almost matches the heat balance constraint $(Q_a|_0^\delta \quad Q_e|_0^r)$ defined in this work. The key result of this study is the clear definition of the additional constraints to the prescribed temperature profiles. This implies, that additionally to the boundary conditions classically imposed to the function approximating the profile by the HBIM two constraints more are necessary: (1) The heat accumulated in the thermal layer (in fact the entire heated body) should be the same either the approximate or the exact temperature profile is used; (2) The heat flux at $x = 0$ defined by the approximate and the exact profile should be the same in the case of PT problem. Similarly, the surface temperature ($x = 0$) defined by either the approximate or the exact temperature profile should be the same in the case of PF problems.

The principle step in solving the problems in the exercises performed here is the formation of additional set of equations allowing to define the parabola exponent and the dimensionless ratio $\delta/(\alpha t)^{1/2}$. In same cases the thermal layer depth equation provided by the HBI and the constraint at $x = 0$ in a particular heat conduction problem are enough to define $n$ and $\delta/(\alpha t)^{1/2}$. However, the introduction of $Q_a|_0^\delta \quad Q_e|_0^r$ as a constraint immediately yields an exponent assuring minimization of the error of approximation. This was clearly demonstrated by the Langford method in the case of prescribed temperature boundary condition problem. The same approach with a suitable change of variables does permit to solve a PT problem for sphere immersed in an infinite medium.

The method developed here defines suitable values of the parabola exponents which are not integer and depend on the type of the boundary condition imposed. The Veinik solution [45], for instance, which is by far different from the Goodman's method, to the PT problem provides $n = 1.79$ within the heat-up period. This, in fact, matches $n = 1.75$ developed in this work and in [17, 24] too. This is not a serious problem in the era of computer added calculations and especially with computer algebra solvers such as Maple and Mathematica. The possibilities to define the parabola exponent through additional constraints enhance the HBIM and gives ground for further research toward solution of more complex problems.

It was demonstrated that when the HBIM and the parabolic profile are applied to a heat-conduction problem with over-specified conditions at $x = 0$ concerning both PT and PF the exponent is definitively determined by the boundary conditions and no additional constraints are needed. Beyond these results, a more general approach automatically generating the deficient condition is developed in the companion article [49] through entropy generation minimization approach.

## Conclusions

The analysis performed in this work clearly elucidates problems emerging in solution through the HBIM when a parabolic profile with an unspecified exponent depending on the type of boundary conditions imposed. The classical Goodman's boundary conditions defining the time-dependent coefficients of the prescribed temperature profile do not work efficiently at the front of the thermal layers if the specific parabolic profile at issue is employed. Additional constraints based on physical assumption enhance the HBIM and form a robust algorithm defining the parabola exponent. The method was demonstrated through detailed solutions of 4 1-D heat-conduction problems in Cartesian co-ordinates.

## Acknowledgments

The help of Prof. O.V. Sharypov (Institute of Thermophysics, RAN, Novosibirsk, Russia) to find the book of Veinik, Prof. V. Zimparov (Technical University, Gabrovo, Bulgaria), and Dr. S. K. Sahu (Indian Institute of Technology, Kharagpur) in discovering early publications on HBIM is highly appreciated. The comments of Prof. T. Myers (University of Cape Town, South Africa), Prof. V. Novozhilov (University of Ulster, UK), Dr. W. Braga (Federal University of Santa Catarina Florianópolis, Santa Catarina, Brazil), and Prof. Nacer Sadoun (Université des sciences et de la technologie Houari Boumediene, Alger) are highly respected.

## Nomenclature

$A$ – dimensionless coefficient in eq. (3a), [–]
Bi – Biot number (= $\alpha_h h_0/\lambda$), [–]
$a$ – coefficient in the prescribed temperature profile, [K]
$b$ – coefficient in the prescribed temperature profile, [K]
$C_p$ – heat capacity, [Jkg$^{-1}$]
$c$ – coefficient in the prescribed temperature profile, [m$^{-1}$]
$E(t)$ – defined integral as a measure of the error of approximations (eq. 62)
$E_n(t)$ – dimensionless function estimating the error of approximations (eqs. 63 and 64a)
$F(t)$ – surface flux, [Wm$^{-2}$]
Fo – Fourier number defined through the slab thickness as a length scale (=$\alpha t/h_0^2$), [–]
Fo$_f$ – Fourier number defined through with $1/\Phi$ as a length scale (=$a\Phi^2 t$), [–]
$h_0$ – slab thickness – see eq. (3a), [m]
$L$ – latent heat, [Jkg$^{-1}$]
$n$ – exponent in the prescribed temperature profile, [–]
$Q$ – heat flux at the surface of heated body ($x = 0$), [W]
$Q_a$ – heat accumulated in the thermal layer calculated through the approximate temperature profile, [W]
$Q_e$ – heat accumulated in the thermal layer calculated through the exact temperature profile, [W]
$\delta Q$ – a function defined by eq. (67), [W]
$q_a$ – surface heat flux provided by the approximate temperature profile, [Wm$^{-2}$]
$q_e$ – surface heat flux provided by the exact temperature profile, [Wm$^{-2}$]
$r$ – radial co-ordinate, [m]
$r_d$ – droplet radius, [m]
$r_0$ – sphere radius (example 4), [m]
$r_{00}$ – radius of the inner heated surface (eq. 7a)
$S$ – face surface of the heated body ($x = 0$), [m$^2$]

$T$ – temperature, [K]
$T_a$ – temperature defined by the approximate solution, [K]
$T_{av}$ – average temperature over the thermal layer [=1/d $\int_0^\delta T(x,t)dx$], [K]
$T_c$ – temperature at the center of a heated droplet (eq. 5a), [K]
$T_e$ – temperature defined by the exact solution, [K]
$T_i$ – initial wood (timber) temperature (eq. 68), [K]
$T_{gas}$ – temperature of the fluid (gas) surrounding the heated body, [K]
$T_s$ – surface temperature (commonly at $x = 0$, see the context), [K]
$T_V$ – vaporization temperature, [K]
$T_\infty$ – temperature of the undisturbed medium, [K]
$t$ – time, [s]
$t_h$ – time required to end the heat-up period (at $\delta = h_0$), [s]
$U(r, t)$ – variable (see eq.48b) = $rT(r, t)$, [mK]
$U_s, U_\infty$ – boundary values of $U(x, t)$ corresponding to $T_s$ ($r = r_0$) and $T_\infty$ ($r$ $\delta$), [mK]
$x, y$ – co-ordinates, [m]
$\tilde{x}$ – $x/h_0$, [–]

*Greek letters*

$\alpha$ – thermal diffusivity, [m$^2$s$^{-1}$]
$\alpha_h$ – heat transfer coefficient (convection heat transfer), [Wm$^{-2}$K]
$\beta_0, \beta_1, \beta_2$ – coefficients of the approximate profile (eq. 49)
$\lambda$ – thermal conductivity, [Wm$^{-1}$K]
$\Phi$ – radiation flux defined length scale (=$F(t)/[\lambda(T_s - T_\infty)]$, [m$^{-1}$]
$\Phi_0$ – radiation-conduction number (=$[F(t)h_0]/[\lambda(T_s - T_\infty)] = N_0$, [–]
$\Theta$ – dimensionless temperature (see eq 2) [= $(T - T_\infty)/(T_s - T\infty)$], [–]

$\delta$   – thermal layer depth, [m]

$\delta_p$   – dimensionless depth of the thermal layer ($=\delta/L$), [–]

*Subscripts*

$p$   – penetration

*Abbreviations*

HBI   – heat balance integral

HBIM   – heat balance integral method
PF   – prescribed flux
PT   – prescribed temperature

*Special symbols*

  – order of magnitude
$a \sim b$   – a scale to b
  – it follows that
  – can be expresses as

Authors´ affiliations:

*J. Hristov*
Department of Chemical Engineering,
University of Chemical Technology and Metallurgy
1756 Sofia, 8 Kl. Ohridsky Blvd., Bulgaria
E-mail: jordan.hristov@mail.bg